\documentclass[onecollarge,natbib]{svjour2}
\bibpunct{[}{]}{;}{n}{}{,} % to get "[numbered]" references from natbib
\smartqed  % flush right qed marks, e.g. at end of proof
\usepackage{amsmath,amsfonts,amssymb,bm}
\usepackage{graphicx}
\usepackage{color}
\usepackage{dsfont}
\usepackage{soul}
\definecolor{purple}{rgb}{0.5,0,0.5}
\definecolor{blue}{rgb}{0.0,0,0.9}
\usepackage{dcolumn}
\usepackage{epsfig}
\usepackage{graphics}
\usepackage{longtable}

\usepackage[colorlinks=true, pdfstartview=FitV, linkcolor=purple, citecolor= purple, urlcolor=blue]{hyperref}
\newcommand{\pslash}{\mbox{$\not \! p$}}

\journalname{Few-Body Systems}
\begin{document}

%...Hadron Structure, Form Factors, and Parton Content
\title{DSE Perspective on QCD Modeling, Distribution Amplitudes, and Form Factors\thanks{Work supported in part by National Science Foundation Grant No. NSF-PHY-1206187}
}
% General acknowledgments should be placed at the end of the article.
%\subtitle{Do you have a subtitle?\\ If so, write it here}

\titlerunning{DSE Perspective}        % if too long for running head

\author{Peter C. Tandy         
%\andSecond Author %etc.
}

%\authorrunning{Short form of author list} % if too long for running head

\institute{P. Tandy \at
              Department of Physics, Kent State University, Kent, Ohio, 44242 USA\\
              %Tel.: +123-45-678910\\
              %Fax: +123-45-678910\\
              \email{tandy@kent.edu}           %  \\
%             \emph{Present address:} of F. Author  %  if needed
          % \and
           %S. Author \at
              %second address
}

%\date{Received: date / Accepted: date}
\date{April 2, 2014}
% The correct dates will be entered by the editor

\maketitle
\begin{abstract}
We describe results for the pion distribution amplitude (PDA) at the  non-perturbative scale \mbox{$\mu=$}\,2\,GeV by projecting the Poincar\'e-covariant Bethe-Salpeter wave-function onto the light-front and use it to investigate the ultraviolet behavior of the electromagnetic form factor, $F_\pi(Q^2)$, on the entire domain of spacelike $Q^2$.  The significant dilation of this PDA compared to the known asymptotic PDA is a signature of dynamical chiral symmetry breaking (DCSB) on the light front.  
We investigate the transition region of $Q^2$ where non-perturbative behavior of constituent-like quarks gives way to the partonic-like behavior of quantum chromodynamics (QCD). The non-perturbative approach is based on the Dyson-Schwinger equation (DSE) framework for continuum investigations in QCD.    The leading-order, leading-twist perturbative QCD result for $Q^2 F_\pi(Q^2)$ underestimates the new DSE computation by just 15\% on $Q^2\gtrsim 8\,$GeV$^2$, in stark contrast with the result obtained using the asymptotic PDA.  

%up to five keywords.
\end{abstract}

\section{Introduction}
\label{intro:sec1}
%From the first pion PDA paper:
A light front formulation of QCD can translate features that arise purely through the infinitely-many-body nature of relativistic quantum field theory into images whose interpretation is similar to quantum mechanical probability amplitudes. 
A phenomenon for which a quantum mechanical image would be desirable is dynamical chiral symmetry breaking (DCSB).  Strictly impossible in quantum mechanics with a finite number of degrees-of-freedom, this emergent feature of  QCD correlates
numerous aspects of the spectrum and interactions of hadrons; e.g., the large splitting between parity partners \cite{Chang:2011ei,Chen:2012qr} and the existence and location of a zero in some hadron form factors \cite{Wilson:2011aa}.   However DCSB has not yet been realized in the light-front formulation of quantum field theory.

The impact of DCSB is expressed with particular force in properties of the pion--the pseudo-Goldstone boson.   Its very existence as the lightest hadron is grounded in DCSB.   The modern paradigm views the pion as both a conventional bound-state in quantum field theory and the Goldstone mode associated with DCSB in QCD.
Numerous model-independent statements may be made about the pion's Bethe-Salpeter amplitude and its relationship to the dressed-quark propagator \cite{Maris:1998hd}.   
We review here the recent extraction of the pion's light-front valence-quark distribution amplitude (PDA) computed from these two quantities.   The results expose DCSB in a covariant wave-function projected onto the light front.

% From the piFF elastic PRL:
There exist exact results and empirically satisfactory results for both soft and hard scattering processes as described in recent reviews\,\cite{Chang:2011vu,Bashir:2012fs,Cloet:2013jya}.    While  the comparison between low-energy experiments and theory check global symmetries and breaking patterns that can be characteristic of a broad class of theories, the high-energy experiments and calculations are a direct probe of specific details of QCD itself; and there are unresolved issues.   Resolutions for a couple of these  are discussed  here.

\section{Pion Distribution Amplitude}
\label{sec:2}

% from first pi DA PRL
The leading twist 2-particle  PDA is given by\footnote{We use a Euclidean metric:  $\{\gamma_\mu,\gamma_\nu\} = 2\delta_{\mu\nu}$; $\gamma_\mu^\dagger = \gamma_\mu$; $\gamma_5= \gamma_4\gamma_1\gamma_2\gamma_3$, tr$[\gamma_5\gamma_\mu\gamma_\nu\gamma_\rho\gamma_\sigma]=-4 \epsilon_{\mu\nu\rho\sigma}$; $\sigma_{\mu\nu}=(i/2)[\gamma_\mu,\gamma_\nu]$; $a \cdot b = \sum_{i=1}^4 a_i b_i$; and $P_\mu$ timelike $\Rightarrow$ $P^2<0$.}

\begin{equation}
f_\pi\, \varphi_\pi(x;\mu) = Z_2 \, {\rm tr}_{\rm CD}\; \int_{dk}^\Lambda \!\!
\delta(n\cdot k - x \,n\cdot P) \,\gamma_5\gamma\cdot n\, \chi_\pi(k;P)\,,
\label{pionPDA}
\end{equation}
where: $f_\pi$ is the pion's leptonic decay constant; the trace is over color and spinor indices; $\int_{dk}^\Lambda$ is a Poincar\'e-invariant regularization of the four-dimensional integral, with $\Lambda$ the ultraviolet regularization mass-scale; $Z_{2}(\mu,\Lambda)$ is the quark wave-function renormalisation constant, with $\mu$ the renormalisation scale; $n$ is a light-like four-vector, $n^2=0$; $P$ is the pion's four-momentum, $P^2=-m_\pi^2$ and $n\cdot P = -m_\pi$, with $m_\pi$ being the pion's mass; and the pion's Bethe-Salpeter wave-function is
\mbox{$\chi_\pi(k;P) = $} \mbox{$S(k) \Gamma_\pi(q;P) S(k-P)$}
with $\Gamma_\pi$ the Bethe-Salpeter amplitude,  $S$ the dressed light-quark propagator, and $q = k-P/2$ is a convenient relative momentum variable for expression of symmetries.  
As a framework within continuum quantum field theory, the DSE study of Ref.\,\cite{Chang:2013pq} was able to reliably compute arbitrarily many moments $\langle x^m\rangle := \int_0^1 dx \, x^m \varphi_\pi(x)$ of the PDA which 
Eq.\,\eqref{pionPDA} yields in the form
\begin{equation}
 \langle x^m\rangle = \frac{N_c\, Z_2}{f_\pi (n\cdot P)^{m+1} }{\rm tr}_{\rm D}
 \! \int_{dk}^\Lambda \!\!
(n\cdot k)^m \,\gamma_5\gamma\cdot n\, \chi_\pi(k;P)\,.
\label{phimom}
\end{equation}

The dressed-quark propagator, having the general form \mbox{$S(p) =$} \mbox{$ 1/[i \gamma\cdot p \, A(p^2,\mu^2) + B(p^2,\mu^2)]$},
%=\frac{Z(p^2,\zeta^2)}{i\gamma\cdot p + M(p^2)}\,,
%&=&Z(p^2,\zeta^2)/[i\gamma\cdot p + M(p^2)]\,.
can be obtained from the relevant DSE, namely the gap equation \cite{Chang:2011vu,Bashir:2012fs}:
\begin{equation}
S^{-1}(p) = Z_2 \,(i\gamma\cdot p + m^{\rm bm})
+ Z_1 \int^\Lambda_{dq}\!\! g^2 D_{\mu\nu}(p-q)\frac{\lambda^a}{2}\gamma_\mu S(q) \frac{\lambda^a}{2}\Gamma_\nu(q,p) ,
\label{gendseN}
\end{equation}
where: $D_{\mu\nu}$ is the dressed gluon propagator; $\Gamma_\nu$ the quark-gluon vertex; $m^{\rm bm}(\Lambda)$ the current-quark bare mass; and $Z_{1}(\mu,\Lambda)$ the vertex renormalization constant.  We employ the renormalisation procedures of Ref.\,\cite{Maris:1997tm} except that the present work uses the renormalization point, $\mu=2\,$GeV to facilitate comparison with existing lattice-QCD information on a few moments of  $\varphi_\pi(x;\mu)$.
Numerous features of the gap equation, its kernel, and the solution procedure are described in Ref.\,\cite{Maris:1997tm}.
%\paragraph{\bf Pion Amplitude}
The amplitude $\Gamma_\pi$ may be obtained from the Bethe-Salpeter equation, a modern expression of which is explained in Ref.\,\cite{Chang:2009zb}.  The  general form is
\begin{equation}
\Gamma_{\pi}(q;P) = \gamma_5
\left[ i E_{\pi}(q;P) + \gamma\cdot P F_{\pi}(q;P)  
+ \, q\cdot P \gamma\cdot q \, G_{\pi}(q;P) + \sigma_{\mu\nu} q_\mu P_\nu H_{\pi}(q;P) \right], %\rule{0.7em}{0ex}
\label{genGpi}
\end{equation}
where the functions are even in $q\cdot P$.  In the chiral limit, $m_\pi=0$ giving  the Goldberger-Treiman-like identity\mbox{$f_\pi E_\pi(q;0) = $} \mbox{$B(q^2)$}.
This is a pointwise statement of Goldstone's theorem and part of a near complete equivalence between the one-quark problem  and the ground state pseudoscalar two-body problem in QCD.   The gap and Bethe-Salpeter equations are key members of the set of Dyson-Schwinger equations (DSEs), which provide an efficacious tool for the study of hadron properties \cite{Chang:2011vu,Bashir:2012fs,Cloet:2013jya}.
% For one-column wide figures use--------------------------------------------------------------------------------
\begin{figure}
\centering
% Use the relevant command to insert your figure file.
% For example, with the graphicx package use
% 
\includegraphics[width=0.7\linewidth]{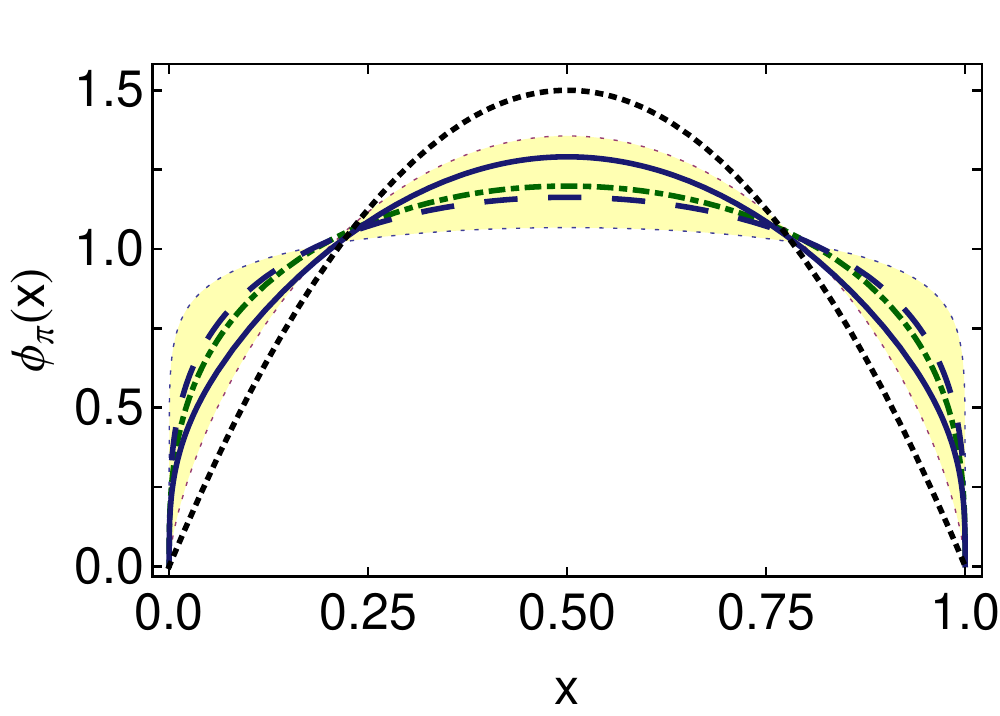}
\caption{ 
$\varphi_\pi(x;\mu)$ at \mbox{$\mu=$} \,2 GeV.  Dot-dashed curve, as determined\,\protect\cite{Cloet:2013tta}  from a single lattice-QCD moment\,\protect\cite{Braun:2006dg} by fitting $\alpha$ according to Eq.\,\protect\eqref{Gegenalpha}.  The shaded region reflects the associated uncertainty. 
The DSE results\,\protect\cite{Chang:2013pq} described in the text are also depicted: solid curve, the DSE-DB result; dashed curve, the DSE-RL result.   The dotted curve is $\varphi_\pi^{\rm asy}(x)$. 
}

\label{fig:phi_RL_DB_latt}     
\end{figure}
%------------------------------------------------------------------------------------------------------------------------------

In Ref.\,\cite{Chang:2013pq}, 50 moments were produced from Eq.\,\eqref{phimom} by employing the interaction of 
Ref.\,\cite{Qin:2011dd} under two different procedures: rainbow-ladder truncation (RL), a very widely used DSE computational scheme in hadron physics, detailed in App.\,A.1 of Ref.\,\cite{Chang:2012cc}; and the DCSB-improved kernel
(DB)\,\cite{Chang:2009zb} detailed in App.\,A.2 of Ref.\,\cite{Chang:2012cc}.  Both schemes are symmetry-preserving, and produce DCSB and Goldstone's theorem, but the DB procedure incorporates  nonperturbative effects associated with DCSB into the kernel itself, and is thus more realistic.   The DB kernel thereby exposes a key role played by the dressed-quark anomalous chromomagnetic moment in determining observable quantities \cite{Chang:2010hb,Chang:2011ei}.   The renormalization  scale \mbox{$\mu = 2 $}~GeV is used to enable direct comparison with lattice-QCD results.    The kernel strength is specified by a product: $D\omega = m_G^3$, and the  effective gluon mass \mbox{$m_G=$}\,0.87\,GeV is the only relevant parameter fitted to properties of ground-state vector and flavor-nonsinglet pseudoscalar mesons\,\cite{Qin:2011dd}.   

The calculation of Eq.\,\eqref{phimom} employed  generalized Nakanishi representations~\cite{Nakanishi:1963zz,Nakanishi:1969ph,Nakanishi:1971BK} for $S(p)$ and $\Gamma_\pi(k;P)$ that are found to provide  excellent parameterizations of previous numerical solutions for $S(k)$ and  $\Gamma_\pi(q;P)$~\cite{Qin:2011dd,Chang:2013pq}.  
The employed representation for the dressed quark propagator is\,\cite{Bhagwat:2003wu}
\begin{equation}
S(p) = \sum_{j=1}^{n_p}  \left(\frac{z_j}{i\pslash + m_j} + \frac{z^\star_j}{i \pslash + m^\star_j}\right)~,
\label{eq:S_masspole}
\end{equation}
where \mbox{$n_p = 2$} pairs of complex conjugate mass poles is found to be sufficient here.   The Nakanishi-type representation of the pion's  Bethe-Salpeter invariant amplitudes ${\cal F}_\sigma$, with \mbox{$\sigma=E, F, G, H$}, can be expressed as
\begin{equation}
{\cal F}_\sigma(q;P) = \int_{-1}^1 d\alpha \, \int_0^\infty d\beta \, \sum_\gamma^{n_t} \frac{\hat{\rho}_\gamma(\alpha,\beta)}
{(q^2 + \alpha q\cdot P + \beta_0 +\beta)^{n_\gamma}}~,
\label{NakBSampl}
\end{equation}
where the $n_\gamma$ are integers, and $\beta_0 > 0$.   It is found that with  \mbox{$n_t = 3$} and 
\mbox{$\hat{\rho}_\gamma(\alpha,\beta) = $} \mbox{$\rho_\gamma(\alpha) \; \delta(\beta + \beta_0 - \Lambda_\gamma^2) $} an excellent representation of numerical solutions is obtained.    For further details and the parameters of 
Eq.\,\eqref{eq:S_masspole} and Eq.\,\eqref{NakBSampl}  see Ref.\,\cite{Chang:2013pq}.  
With these representations the quark loop integral Eq.\,\eqref{phimom} becomes
\begin{equation}
 \langle x^m\rangle \propto \sum_\sigma \; \int_{-1}^1 d\alpha \; \rho_\gamma(\alpha) \;I^m_\sigma(\alpha,n,P)~,
\label{momint}
\end{equation}
where \mbox{$\sigma = {\gamma, j_1, j_2}$} denotes the set of discrete summation labels for the bound state vertex and the two propagators, 
\begin{equation}
I^m_\sigma(\alpha,n,P) = \int \frac{d^4k}{(2\pi)^4} \; \frac{N^m_\sigma(k,n,P)}{ {\cal D}_\sigma(k,P;\alpha) }~,
\label{Feynint}
\end{equation}
and \mbox{$  {\cal D}_\sigma  = [D_\gamma(k,P;\alpha)]^{n_\gamma}\, D_{j_1}(k,P) \, D_{j_2}(k) \, R(k) $}.   
With \mbox{$ R(k) = (1 + k^2/\Lambda^2)$} being the ultraviolet regulator, each denominator factor of the integrand for 
$ I^m_\sigma$ is a quadratic form in $k$ and the numerator $ N^m_\sigma$ contains only limited finite powers, and both are Lorentz invariant.   Use of the standard techniques associated with the Feynman integrals familiar from perturbation theory then yields the final result in algebraic form.  
% For one-column wide figures use--------------------------------------------------------------------------------
\begin{figure}
\centering
% Use the relevant command to insert your figure file.
% For example, with the graphicx package use
% 
\includegraphics[width=0.7\linewidth]{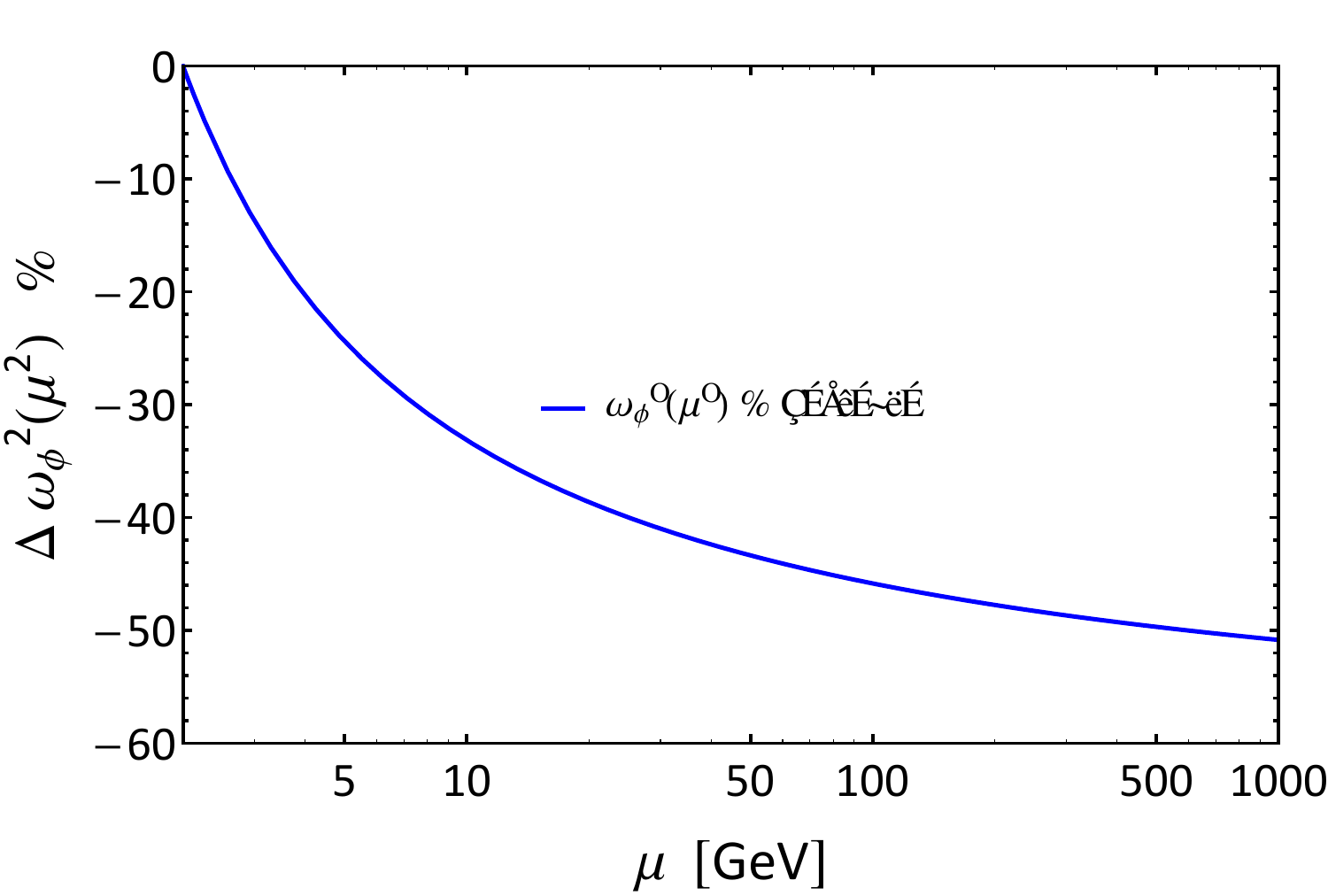}
\caption{ 
The \% change in $\omega_\phi^2(\mu) $ relative to its value at \mbox{$\mu=$}\,2 GeV as it is evolved to higher scales.
}

\label{fig:omega_phi_2}     
\end{figure}
%------------------------------------------------------------------------------------------------------------------------------

The dressed nature of propagators and vertices induces a departure from power law behavior in the ultraviolet due to the  
one-loop renormalization group behavior of QCD as produced  by the particular DSE truncation employed.     For the Nakanishi-type representations this requires incorporation of an additional ultraviolet momentum dependence  like  $\ln^{ -\gamma_F }[p^2/\Lambda_{\rm QCD}^2]$,  where $\gamma_F$ is the anomalous dimension of object $F$ in the loop integral.   A fractional power approximation to the logarithm is found to be quite accurate for the ultraviolet 
behavior\,\cite{Chang:2013nia}; the algebraic results for the relevant Feynman integrals are straightforward.

This technique~\cite{Chang:2013pq}  greatly simplifies the practical problem of continuing from Euclidean  to 
Minkowski metric; it allows the use of Euclidean DSE modeling of QCD to address quantities that are naturally defined by 
light-front momenta.  Accumulated experience in Euclidean QCD-modeling has found that, with the momentum dependence of the gap and Bethe-Salpeter equation solutions represented only by numerical arrays,  the variable mass scales $>> \Lambda_{\rm QCD}$  encountered in  treatments of hadronic observables requires a near impossible accounting for singularities in the complex $p^2$ plane of integration.     The above method will only produce singularities in the Feynman integral result of a given process amplitude if there are observable  production channels and thresholds open to that process.

\subsection{Representations and comparison with lattice-QCD}
%\label{sec:2a}
% (see Sect.~\ref{sec:1}).

%\paragraph{\bf Reconstruct DSE $\phi$, Gegen-$\alpha$}  
From Eq.\,\eqref{phimom} the PDA moments $\{\langle x^m\rangle|m=1,\ldots,m_{\rm max}\}$ with $m_{\rm max}=50$ are used to reconstruct $\phi_\pi(x;\mu)$.
Since the Gegenbauer polynomials  $C_n^{(\alpha+1/2)}(2 x -1)$ are a complete orthonormal set on $x\in[0,1]$ with respect to the measure $[x (1-x)]^{\alpha}$,  they facilitate reconstruction of any function that vanishes at $x=0,1$ and is symmetric about $x=1/2$.  One therefore fits the $\langle x^m \rangle$ to the moments of 
\begin{equation}
\varphi_\pi(x;\mu) = N_\alpha \, [x (1-x)]^{\alpha}\;
\bigg[ 1 + \sum_{n=2,4,\ldots}^{n_{max}} a_n^\alpha(\mu) \;C_n^{(\alpha+1/2)}(2 x -1) \bigg],
\label{Gegenalpha}
\end{equation}
where $N_\alpha = \Gamma(2\alpha+2)/[\Gamma(\alpha+1)]^2$.  
The value of $\alpha$ can be optimized to minimize the number of terms ($n_{max}$) needed to fit the $\langle x^m \rangle$, thus  producing a rapidly convergent series.   A value \mbox{$n_{max}=2 $} is then found to ensure that the $\langle x^m \rangle$ are reproduced to within an RMS error of 1\%.  In general, the quantities $\alpha$, and $n_{max}$ are also dependent on the scale $\mu$.   The dashed curve in Fig.~\ref{fig:phi_RL_DB_latt} is the RL result for the PDA at 2~GeV.   It is described by Eq.\,\eqref{Gegenalpha} with $\alpha_{\rm RL} = 0.29$,  and the only coefficient needed is $a_2^{\rm RL}=0.0029$.  The solid curve in Fig.\,\ref{fig:phi_RL_DB_latt} is the DB result and it is described by Eq.\,\eqref{Gegenalpha} with $\alpha_{\rm DB} = 0.31$,  and the only coefficient needed is $a_2^{\rm DB}=-0.12$.  Compared to the asymptotic QCD result \mbox{$\phi_\pi^{\rm asy}=$} \mbox{$6x(1-x) $}, the non-perturbative PDA in Fig.\,\ref{fig:phi_RL_DB_latt}  is significantly dilated.  This can be traced to DCSB and is a long-sought and unambiguous expression of that phenomenon on the light-front \cite{Brodsky:2010xf,Chang:2011mu,Brodsky:2012ku}.

If one were to instead impose  \mbox{$\alpha=1 $}, then the representation in Eq.\,\eqref{Gegenalpha} becomes identical to the familiar perturbative QCD representation in terms of the Gegenbauer $C^{(3/2)}_n(2x-1)$ polynomials which are the irreducible representations of the  collinear conformal group SL$(2;\mathbb{R})$ that expresses the invariance of QCD at asymptotically large scales~\cite{Efremov:1979qk,Lepage:1980fj,Brodsky:1980ny,Braun:2003rp}.  
%SL$(2;\mathbb{\rule{0.2ex}{1.5ex}\rule{0.1ex}{0ex}R})$
The scale evolution of the coefficients $a^{3/2}_n(\mu)$ is known~\cite{Efremov:1979qk,Lepage:1980fj}
 and is especially simple at leading order.   Because of the certainty of asymptotic QCD results, and  the absence of  information on $ \varphi_\pi(x;\mu)$ at a non-perturbative $\mu$, it has been common to seek a determination of the first few coefficients  $a^{3/2}_n(\mu)$ with limited information from lattice-QCD, high energy exclusive scattering or the  QCD sum rule approach, even though the scale would be finite\,\cite{Mikhailov:2014rqa,Braun:2006dg}.  
However, the PDA extracted from the DSE work at \mbox{$\mu = 2 $}~GeV, when projected onto a $\{C_n^{(3/2)}\}$-basis, shows that many more than a few coefficients $\{a^{3/2}_n\}$ are needed.       For both DSE results, one needs  \mbox{$n_{max} > 14 $} before \mbox{$a_n^{3/2} < 0.1\, a_2^{3/2} $} and the PDA is adequately reproduced.   A truncation to  \mbox{$n_{max}=4 $}  introduces spurious oscillations, or multiple-humped PDAs,  that are typical of non-converged Fourier-like representations.  Since the pion multiplet contains a charge-conjugation eigenstate, each of the three significant invariant amplitudes of  $\Gamma_\pi(q;P)$ peaks at  zero relative momentum $q^2$ and monotonically decrease with $q^2$, as confirmed by solution of the Bethe-Salpeter equation.    As a consequence  $\varphi_\pi(x;\mu)$ should exhibit a single maximum at $x=1/2$.   In seeking to extract $\varphi_\pi(x)$ from limited information,  it is better to fit $\alpha$ first than to force $\alpha=1$ and infer a value for a few of the many sizable $a_n^{(3/2)}$.   
 
That non-perturbative scales entail large corrections to $\varphi^{\rm asy}_\pi(x)$ is illustrated by  the scale evolution of the $a^{3/2}_n(\mu)$ obtained by projection of the DSE PDA.    After evolution to $\mu=100\,{\rm GeV}$,  $a_2^{3/2}(\mu)$ has fallen to only 50\% of its 2~GeV value, while $a_4^{3/2}(\mu)$ has fallen to only 37\% of its 2~GeV value.   These observations suggest that asymptotic QCD is quite remote from present experimental capabilities. 
 
%---from 'PDA from Lattice" PRL:  %Ref.\,\cite{Cloet:2013tta}
%\paragraph{\bf Compare to lattice moms and distn}   
The PDA moment result \mbox{$\langle (2x-1)^2 \rangle_{\phi_\pi} =$} \mbox{$ 0.27\pm 0.04$} was obtained from lattice-QCD\,\cite{Braun:2006dg}.
%\int_0^1 dx\, (2 x - 1)^2 \, \varphi_\pi(x,\tau_2) = 0.27\pm 0.04\,.
%0:269(39)
With the conventional Gegenbauer-(3/2) version of Eq.\,\eqref{Gegenalpha}  used to determine the single coefficient  $a_2^{3/2}(\mu_2)$, a ``double-humped'' PDA was produced\,\cite{Braun:2006dg}.  
If instead one views $\alpha$ in  Eq.\,\eqref{Gegenalpha} as the single parameter to be determined, the result is \mbox{$\alpha = $}
%\mbox{$0.35^{+ 0.32 = 0.67}_{-0.24=0.11} $}
\mbox{$0.35^{+ 0.32}_{-0.24} $}\,\cite{Cloet:2013tta}.   This compares favorably with the values 0.31 and 0.29 obtained from the two DSE kernels~\cite{Chang:2013pq} , and again indicates that very few expansion coefficients $a^\alpha_n$ would be needed to improve the PDA.
The PDA determined this way from lattice-QCD is depicted in Fig.\,\ref{fig:phi_RL_DB_latt} along with the band reflecting lattice  uncertainties.  It  produces a concave amplitude in agreement with contemporary DSE studies and confirms that  $\varphi_\pi^{\rm asy}(x)$ is not a good approximation at $\mu=2\,$GeV, and an expansion that starts with it will have poor convergence properties.
Projection of this DSE-inspired fit onto the Gegenbauer-(3/2) basis allows the scale to be evolved for application to a given process.
One finds that  only for $\mu\gtrsim 100\,$GeV is $a_2^{3/2}\lesssim 10$\% and $a_4^{3/2}/a_2^{3/2}\lesssim 30$\%.  Evidently, the influence of DCSB, which causes the PDA to be broader than the asymptotic QCD result, persists to remarkably high scales.

\section{Pion Charge Form Factor}
\label{sec:3}

%----------------------------------------------------------------------------------------------------------------------------
%  one-column  figure
\begin{figure}
\centering
% Use the relevant command to insert your figure file.
% For example, with the graphicx package use
% figure caption is below the figure
\includegraphics[width=0.67\linewidth]{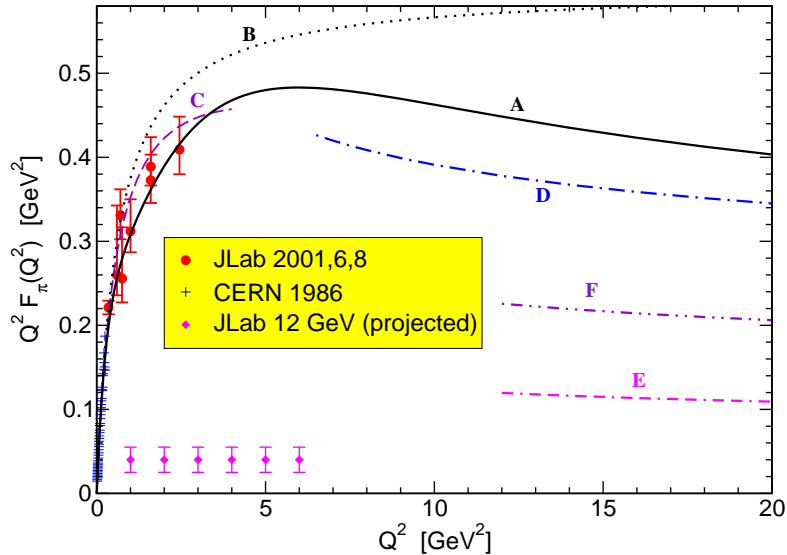}
\caption{$Q^2 F_\pi(Q^2)$.  Solid black curve (A):  the 2013 DSE result\,\protect\cite{Chang:2013nia};  dotted black curve (B): the vector meson dominance Ansatz using the $\rho$ mass; long dashed purple curve (C): the previous 2000 DSE result\,\protect\cite{Maris:2000sk} that could not address \mbox{$Q^2 > $}\,4\,GeV$^2$; dash-dot blue curve (D): the uv-QCD result of Eq.\,\protect\eqref{pionUV} using $\varphi_\pi(x;\mu)$ at 2\,GeV;  dash-dash-dot magenta curve (E): the asymptotic or conformal QCD result using $\varphi^{\rm asy}_\pi(x) $;  dash-dot-dot purple curve (F): the uv-QCD result  using $\varphi_\pi(x;\mu)$ at 10\,GeV.   The filled red circles are the data described in Ref.\,\protect\cite{Huber:2008id}; and the filled diamonds indicate the projected reach and accuracy of a forthcoming experiment\,\protect\cite{E1206101}.   The short horizontal lines at very low $Q^2$ depict the data from Ref.\,\protect\cite{Amendolia:1986wj}.
}
\label{fig:Q2piFF2013}      
\end{figure}
%--------------------------------------------------------------------------------------------------------------------------

The QCD prediction for exclusive scattering at large $Q^2$ follows from observations that the process amplitude factorizes into a perturbative scattering amplitude  that supports the flow of hard momentum convoluted with  a soft amplitude carrying the non-perturbative dynamics of the initial and final hadron state\,\cite{Farrar:1979aw,Efremov:1979qk,Lepage:1980fj}.   With light front coordinates being an efficient way to represent this, hadronic distribution amplitudes enter into the description.
The ultraviolet behavior of $F_\pi(Q^2)$ is of great contemporary interest.  The rainbow-ladder DSE prediction\,\cite{Maris:2000sk} in 2000 for $Q^2 F_\pi(Q^2)$ in Fig.\,\ref{fig:Q2piFF2013_DynConsM} agrees with the existing accurate data but it only hints at a maximum at $Q^2\approx 6\,$GeV$^2$.  The domain upon which this quantity flattens is expected be accessible to next-generation experiments \cite{E1206101}.  The QCD prediction for the pion charge form factor at suitably large $Q^2$\,\cite{Farrar:1979aw,Efremov:1979qk,Lepage:1980fj} is
\begin{equation}
\label{pionUV}
Q^2 >> \Lambda_{\rm QCD}^2 : \;  Q^2 F_\pi(Q^2) \to 16 \pi \alpha_s(Q^2)  f_\pi^2 \; \omega_\phi^2(Q^2) + \mathcal{O}(1/Q^2)\,,
\end{equation}
where $f_\pi=92.2\,$MeV is the pion decay constant, and
\begin{equation}
\label{wphi}
\omega_\phi(\mu^2) = \frac{1}{3} \int_0^1 dx\, \frac{1}{x} \varphi_\pi(x;\mu)\,.
\end{equation}
%where $\varphi_\pi(x;\mu)$ is the pion's valence-quark parton distribution amplitude (PDA) at scale $\mu$.  
For the purposes of Eq.\,\eqref{pionUV} the choice \mbox{$\mu = Q$} is representative.  If one uses the asymptotic PDA
\mbox{$\varphi_\pi(x;\infty) = $} \mbox{$\varphi^{\rm asy}_\pi(x) $} then \mbox{$\omega_\phi \to1$},
and Eq.\,\eqref{pionUV} reduces to the well-used asymptotic expression for $ Q^2 F_\pi(Q^2$~\cite{Farrar:1979aw,Efremov:1979qk,Lepage:1980fj}.    Just how large $Q^2$ must be  for it to be accurate has not been clear.  

At $Q^2=4\,$GeV$^2$, approximately the midpoint of the domain accessible at next-generation facilities, the asymptotic expression yields \mbox{$ Q^2 F_\pi(Q^2=4) = 0.15$} with $n_f=4$ and $\Lambda_{\rm QCD}=0.234\,$GeV~\cite{Qin:2011dd}.  This  is a factor of $2.7$ smaller than the empirical value ($0.41^{+0.04}_{-0.03}$) quoted at $Q^2 =2.45\,$GeV$^2$ \cite{Horn:2006tm,Huber:2008id}, and a factor of three smaller than the previous DSE theory result at $Q^2 =4\,$GeV$^2$ in Ref.\,\cite{Maris:2000sk}.  Notably, Ref.\,\cite{Maris:2000sk} provided the only prediction for the pointwise behavior of $F_\pi(Q^2)$ that is applicable on the entire spacelike domain currently mapped reliably by experiment.  It seemed a reasonable assumption that, by about \mbox{$Q^2 \gtrsim $}\,10\,GeV$^2$ perturbative scattering mechanisms and partonic behavior should have set in, so that after removal of the valence quark counting power, only the slow logarithms of QCD remain.   It was therefore difficult to imagine that the magnitude 
of $ Q^2 F_\pi(Q^2)$ will fall by a factor of $3$ as $Q^2$  covered the range  4 to 10 GeV$^2$.

Two recent developments have cleared up much of this matter.  Firstly, the recent DSE calculation\,\cite{Chang:2013pq}  of $\varphi_\pi(x;\mu)$ provides information about $\omega_\phi(\mu^2) $  for Eq.\,\eqref{pionUV} in  this $Q^2$ range.   Secondly, the DSE approach to $F_\pi(Q^2)$ has been reformulated  with new methods that enable a calculation\,\cite{Chang:2013nia} to arbitrarily large-$Q^2$, thus allowing a consistent examination of the transition between non-perturbative and perturbative QCD regimes.  
%\paragraph{\bf The DSE results for $\omega_\phi(\mu^2)$}   
It is the quantity $\omega_\phi^2(Q^2)$ that describes how the perturbative QCD domain links to the asymptotic domain.   The DSE results for $\omega_\phi(\mu^2)$ can be analyzed in terms of the equivalent number of $C_n^{(3/2)}(2 x -1)$ polynomials needed if that representation is chosen for a range of $\mu$ in the ultraviolet.  To reproduce more than 90\% of $\omega_\phi^2(\mu)$,  the number  needed is 9 at $\mu = 2$~GeV, 7 at $\mu = 4$~GeV, and 5 at $\mu = 10$~GeV.    It is therefore necessary to build $\omega_\phi^2(\mu^2)$ from a successful description of pion structure at non-perturbative scales.   The scale evolution of the projected  coefficients $a_n^{3/2}(\mu)$, and thus  $\omega_\phi^2(\mu^2) $, away from the non-perturbative region is very slow.    In  Fig.~\ref{fig:omega_phi_2} is displayed the \% decrease of $\omega_\phi^2(\mu^2) $ with increasing scale, relative to its value at \mbox{$\mu=$}\,2\,GeV.   Even at 1\,TeV it still holds about 50\% of its value associated with the non-perturbative domain.  

\subsection{Pion Form Factor Calculation}
%\label{sec:2a}
% (see Sect.~\ref{sec:1}).
%\paragraph{\bf Pion Form Factor Calculation} 
At leading order in the  symmetry-preserving rainbow-ladder DSE truncation scheme reviewed in Refs.\,\cite{Chang:2011vu,Bashir:2012fs}, the pion form factor is given by
\begin{equation}
2\,K_\mu F_\pi(Q^2)  =  N_c {\rm tr}_{\rm D} %\int_{dk}^\Lambda
\int\! \frac{d^4 k}{(2\pi)^4}\, \chi_\mu(k+p_f,k+p_i) 
 \Gamma_\pi(k_i;p_i)\,S(k)\,\Gamma_\pi(k_f;-p_f)\,, 
\label{RLFpi}
\end{equation}
where $Q$ is the incoming photon momentum, $p_{f,i} = K\pm Q/2$, $k_{f,i}=k+p_{f,i}/2$, and the remaining trace is over spinor indices. The unamputated dressed-quark-photon vertex, $\chi_\mu(k_f,k_i)$, should also be computed in RL truncation.   The impact of corrections to this truncation is understood and the dominant effect is a modified power associated with the logarithmic running; in either case the running is so slow that the omitted diagrams have no material impact here.    
%----------------------------------------------------------------------------------------------------------------------------
%  one-column  figure
\begin{figure}
\centering
% Use the relevant command to insert your figure file.
% For example, with the graphicx package use
% figure caption is below the figure
\includegraphics[width=0.74\linewidth]{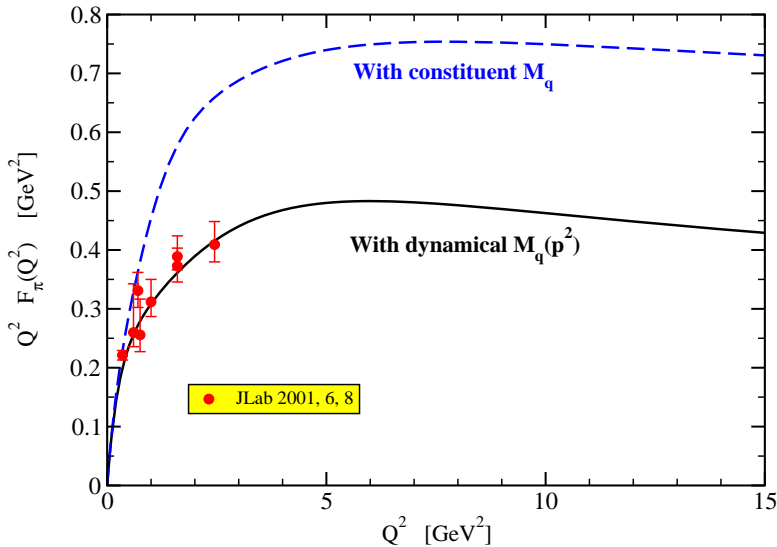}
\caption{Dynamical mass effect.  Solid curve (black): the 2013 DSE result\,\protect\cite{Chang:2013nia} containing the momentum-dependent quark mass dressing; dashed curve (blue): the quark propagators external to the bound state vertices are replaced by constituent mass propagators.  The dynamical mass yields a reduction of about 64\% near the turnover.    
}
\label{fig:Q2piFF2013_DynConsM}      
\end{figure}
%--------------------------------------------------------------------------------------------------------------------------

In the most recent work\,\cite{Chang:2013nia}, we employ the RL interaction of Ref.\,\cite{Qin:2011dd} and  the generalized Nakanishi spectral representations~\cite{Nakanishi:1963zz,Nakanishi:1969ph,Nakanishi:1971BK} for $S(p)$ and $\Gamma_\pi(k;P)$ as described in association with Eqs.\,\eqref{eq:S_masspole} and\,\eqref{NakBSampl}.      In a straightforward generalization of the analysis presented earlier for the pion distribution amplitude in Eqs.\,\eqref{momint} and\,\eqref{Feynint}, the calculation  of $F_\pi(Q^2)$ reduces to a sum of standard Feynman integrals, and the result is an algebraic expression.
For the unamputated dressed-quark-photon vertex, $\chi_\mu(k_f,k_i)$, we use an \emph{Ansatz}  with the following properties.   It satisfies the longitudinal Ward-Green-Takahashi identity, 
%\,\cite{Ward:1950xp,Green:1953te,Takahashi:1957xn}
is free of kinematic singularities, reduces to the bare vertex in the free-field limit, and has the same Poincar\'e transformation properties as the bare vertex.  The \emph{Ansatz} also includes a dressed-quark anomalous magnetic moment, made mandatory by DCSB \cite{Chang:2010hb}.
%\cite{Bashir:2011dp}
This and other non-perturbative corrections to the bare vertex, including the tail of the $\rho$-meson resonance, are negligible for spacelike momenta $Q^2\gtrsim 1\,$GeV$^2$\,\cite{Alkofer:1993gu,Maris:1999bh}.  
% $\rho$ pole has only a modest impact on $Q^2 r_\pi^2 \lesssim 1$, where $r_\pi$ is the pion's charge radius

The new DSE calculation\,\cite{Chang:2013nia} of $F_\pi(Q^2)$ is displayed in Fig.\,\ref{fig:Q2piFF2013}.  The dash-dash-dot curve is the prediction of QCD at a truely asymptotic scale, and uses 
$\varphi_\pi^{\rm asy}(x)$.     It is often characterized as the prediction of pQCD and thus relevant to the range  $Q^2\gtrsim $6\,GeV$^2$ in such plots.   However as discussed earlier  $\varphi_\pi(x;\mu)$ is significantly broader than $\varphi_\pi^{\rm asy}(x)$ at the displayed scales, and this makes $\omega_\phi(\mu^2)$ almost a factor of 3 larger than its asymptotic value 1.0.   To be more specific, with $\varphi_\pi(x;\mu)$ taken at \mbox{$\mu = $}\,2\,GeV, the uv-QCD prediction from  Eq.\,\eqref{pionUV} is shown by the dash-dot curve; this is only 15\% below the 
most recent DSE calculation which is the solid curve\,\cite{Chang:2013nia}.     With $\varphi_\pi(x;\mu)$ taken at \mbox{$\mu = $}\,10\,GeV, the uv-QCD prediction is depicted by the dash-dot-dot curve, which is really only relevant at \mbox{$Q^2=$} \, 100\,GeV$^2$ but shown here to indicate how slow is the approach to the asymptotic result\footnote{A continuous scale evolution of 
$\varphi_\pi(x;Q)$ is ignored in Fig.\,\ref{fig:Q2piFF2013_DynConsM}  because it is too slow to be depicted properly on this limited $Q^2$ domain}.
The findings can be summarized by  $\omega_\phi^2(4\,{\rm GeV}^2) = 3.2$,  and $\omega_\phi^2(100\,{\rm GeV}^2) = 2.0$, with its evolution over a wide range of scales shown in Fig.~\ref{fig:omega_phi_2}.  Even at 1\,TeV it still holds about 50\% of its value associated with scales accessible to experiment.  

In Fig.\,\ref{fig:Q2piFF2013_DynConsM} we display the strong reduction effect that the dynamical dressing of the quark mass function has on the ultraviolet magnitude of $ Q^2 F_\pi(Q^2)$.   The dashed curve is the result obtained if the quark propagators evident in the triangle diagram are replaced by constituent mass propagators.  The full calculation, shown by the solid curve, naturally includes the momentum-dependent evolution of the quark mass from a typical constituent value at quark momenta 
\mbox{$k^2 \lesssim 1$}\,GeV$^2$ to its small partonic value for \mbox{$k^2 \gtrsim 4$}\,GeV$^2$.   With  \mbox{$k^2 \sim 2$}\,GeV$^2$ taken as a measure of the transition region, and noting that as $Q^2$ increases there is a fixed spacelike bias of $Q/2$ in each   
quark propagator, one estimates that \mbox{$Q^2 \sim $}\,8\,GeV$^2$ is a good measure of the IR-UV transition or turnover region\,\cite{Maris:1998hc}.  There the dynamical mass yields a reduction of about 64\%.   
%----PCT--end from piFF paper

\section{Summary}
\label{sec:4}
%\paragraph{\bf Summary} 
$\varphi_\pi^{\rm asy}(x)$ is a poor approximation to $\varphi_\pi(x;\mu)$ at all momentum-transfer scales that are either now accessible to experiments involving pion elastic or transition processes, or will become so in the foreseeable future\,\cite{Holt:2012gg,Dudek:2012vr}.
%\cite{Huber:2008id,Uehara:2012ag,Holt:2012gg,Dudek:2012vr} 
Predictions of leading-order, leading-twist formulae involving $\varphi^{\rm asy}_\pi(x)$ are a misleading guide to interpreting and understanding contemporary experiments.  At accessible energy scales a better guide is obtained by using the broad PDA described herein in such formulae.  This might be adequate for the charged pion's elastic form factor.  However, it will probably be necessary to consider higher twist and higher-order corrections in controversial cases such as the $\gamma^\ast \gamma \to \pi^0$ transition form factor \cite{Roberts:2010rn,Brodsky:2011yv,Bakulev:2012nh}.
%also \cite{Agaev:2010aq}

The near agreement for \mbox{$Q^2 > 6$}\,GeV$^2$ in Fig.\,\ref{fig:Q2piFF2013} between the perturbative QCD prediction that uses $\varphi_\pi(x;\,2\,{\rm GeV})$ (dash-dot curve) and the new DSE result for $Q^2 F_\pi(Q^2)$ (solid curve) is striking.  It highlights that a single DSE interaction kernel, essentially determined by one strength parameter, and preserving the one-loop renormalization group behavior of QCD, is very close to unifying the pion's electromagnetic form factor and its valence-quark distribution amplitude.  Numerous other quantities are also correlated quite closely via a single DSE interaction kernel \,\cite{Chang:2011vu,Bashir:2012fs,Cloet:2013jya,Eichmann:2011ej}.

Moreover, this leading-order, leading-twist QCD prediction, obtained with a pion valence-quark PDA evaluated at a scale appropriate to the experiment, underestimates our full computation by merely an approximately uniform 15\% on the domain depicted.  The small mismatch should be  explained by a combination of higher-order, higher-twist corrections to Eq.\,\eqref{pionUV} and shortcomings in the rainbow-ladder truncation, which predicts the correct power-law behavior for the form factor but not precisely the right anomalous dimension.  Hence, as anticipated earlier \cite{Maris:1998hc} (and expressing a result that can be understood via the behavior of the dressed-quark mass-function \cite{Chang:2011vu,Bashir:2012fs}), one should expect dominance of hard contributions to the pion form factor for $Q^2\gtrsim 8\,$GeV$^2$.  Notwithstanding this, the normalization of the form factor is fixed by a pion wave-function whose dilation with respect to $\varphi_\pi^{\rm asy}(x)$ is a definitive signature of dynamical chiral symmetry breaking.

\begin{acknowledgements}
I wish to acknowledge valuable interactions with C.~D.~Roberts and Lei~Chang that made a lot of this work possible. I also wish to thank the organizers of the LightCone 2013 workshop for providing a fine program and a welcoming atmosphere.   This work was supported in part by the National Science Foundation under Grant No. NSF-PHY-1206187.
\end{acknowledgements}

% BibTeX users please use
%%%%%\bibliographystyle{spbasic}

%\bibliographystyle{spbasic4FBS}
%\bibliography{refsPM,refsPCT,refsCDR,refs,refsMAP}   % name your BibTeX data base
% now input the .bbl file with end comments removed and any problems edited by hand!

\end{document}